\documentclass[structabstract]{aa}  
\usepackage[hyperindex=true,colorlinks=true,citecolor=blue,linkcolor=blue,breaklinks=true]{hyperref}
\usepackage{amsmath}
\usepackage{graphicx}
\usepackage{txfonts}
\usepackage{natbib,twoopt}
\usepackage{color}

\bibpunct{(}{)}{;}{a}{}{,} %% natbib format like A&A and ApJ 
\newcommandtwoopt{\citeads}[3][][]{\href{http://adsabs.harvard.edu/abs/#3}% 
{\citealp[#1][#2]{#3}}} 
\newcommandtwoopt{\citepads}[3][][]{\href{http://adsabs.harvard.edu/abs/#3}% 
{\citep[#1][#2]{#3}}} 
\newcommandtwoopt{\citetads}[3][][]{\href{http://adsabs.harvard.edu/abs/#3}% 
{\citet[#1][#2]{#3}}}
\newcommandtwoopt{\citeyearads}[3][][]% 
{\href{http://adsabs.harvard.edu/abs/#3}{\citeyear[#1][#2]{#3}}}

\begin{document}

	\title{Multiplicity of Galactic Cepheids from long-baseline interferometry}
	\titlerunning{Multiplicity of Galactic Cepheids from long-baseline interferometry}

   	\subtitle{I.~CHARA/MIRC detection of the companion of V1334~Cygni}

	\author{ A.~Gallenne\inst{1}, 
				J.~D.~Monnier\inst{2},
				A.~M\'erand\inst{3},
				P.~Kervella\inst{4}, 
				S.~Kraus\inst{5},
				G.~H.~Schaefer\inst{6},  
				W.~Gieren\inst{1}, 
				G.~Pietrzy\'nski\inst{1,9},
				L.~Szabados\inst{7},
				X.~Che\inst{2},
				F.~Baron\inst{2},
				E.~Pedretti\inst{8},
				H.~McAlister\inst{6}, T.~ten~Brummelaar\inst{6}, J.~Sturmann\inst{6}, L.~Sturmann\inst{6}, N.~Turner\inst{6}, C.~Farrington\inst{6} \and N.~Vargas\inst{6} 			
  				}
  				
  	\authorrunning{A. Gallenne et al.}

\institute{Universidad de Concepci\'on, Departamento de Astronom\'ia, Casilla 160-C, Concepci\'on, Chile
	\and Astronomy Department, University of Michigan, 1034 Dennison Bldg, Ann Arbor, MI 48109-1090, USA
	\and European Southern Observatory, Alonso de C\'ordova 3107, Casilla 19001, Santiago 19, Chile
	\and LESIA, Observatoire de Paris, CNRS UMR 8109, UPMC, Universit\'e Paris Diderot, 5 Place Jules Janssen, F-92195 Meudon, France
	\and Harvard-Smithsonian Center for Astrophysics, 60 Garden Street, MS-78, Cambridge, MA 02138, USA
  	\and The CHARA Array of Georgia State University, Mount Wilson CA 91023, USA
  	\and Konkoly Observatory, Research Centre for Astronomy and Earth Sciences, Hungarian Academy of Sciences, H-1121 Budapest, Konkoly Thege Mikl\'os \'ut 15-17, Hungary
  	\and School of Physics and Astronomy, University of St Andrews, North Haugh, St Andrews, Fife, KY16 9SS, UK
  	\and Warsaw University Observatory, Al. Ujazdowskie 4, 00-478, Warsaw, Poland}%\\
  
  \offprints{A. Gallenne} \mail{agallenne@astro-udec.cl}

   \date{Received January 11, 2013; accepted February 6, 2013}

% \abstract{}{}{}{}{} 
% 5 {} token are mandatory
 
  \abstract
   {More than 60\,\% of Cepheids are in binary or multiple systems. Studying such systems could lead to a better understanding of the age and evolution of Cepheids. These are also useful tools to estimate the mass of Cepheids, and constrain theoretical models of their pulsation and evolution.}
   {We aim at determining the masses of Cepheids in binary systems, as well as their geometric distances and the flux contribution of the companions. The combination of interferometry with spectroscopy will offer a unique and independent estimate of the Cepheid masses.}
   {Using long-baseline interferometry at visible and infrared wavelengths, it is possible to spatially resolve binary systems containing a Cepheid down to milliarcsecond separations.  Based on the resulting visual orbit and radial velocities, we can then derive the fundamental parameters of these systems, particularly the masses of the components and the geometric distance. We therefore performed interferometric observations of the first-overtone mode Cepheid V1334~Cyg with the CHARA/MIRC combiner.}
   {We report the first detection of a Cepheid companion using long-baseline interferometry. We detect the signature of a companion orbiting V1334~Cyg at two epochs. We measure a flux ratio between the companion and the Cepheid $f = 3.10 \pm 0.08$\,\%, giving an apparent magnitude $m_\mathrm{H} = 8.47 \pm 0.15$\,mag. The combination of interferometric and spectroscopic data have enabled the unique determination of the orbital elements.
   $P = 1938.6 \pm 1.2$\, days, $T_\mathrm{p} = 2~443~616.1 \pm 7.3$, $a = 8.54 \pm 0.51$\,mas, $i = 124.7 \pm 1.8\degr$, $e = 0.190 \pm 0.013$, $\omega = 228.7 \pm 1.6\degr$, and $\Omega = 206.3 \pm 9.4\degr$. We derive a minimal distance $d \sim 691$\,pc, a minimum mass for both stars of $3.6\,M_\odot$, with a spectral type earlier than B5.5V for the companion star. Our measured flux ratio suggests that radial velocity detection of the companion using spectroscopy is within reach, and would provide an orbital parallax and model-free masses.}
   {}

 
 \maketitle

%
%________________________________________________________________

\section{Introduction}
Classical Cepheid stars have been considered as reliable tools to estimate distances in the universe for more than a century \citep[see e.g.][]{Leavitt_1912_03_1,Fernie_1969_12_0,Sandage_2006_09_0,Barnes_2009_09_0,Bono_2010_05_0}. Their Period-Luminosity (P--L) relation makes them valuable to determine extragalactic distances and to calibrate secondary distance indicators. In addition to the determination of distances, Cepheids are also powerful astrophysical laboratories that provide fundamental clues for studying the pulsation and evolution of intermediate-mass stars \citep[see e.g.][]{Prada-Moroni_2012_04_0,Bono_2006__0,Caputo_2005_08_0}.

The occurrence of Cepheids in binary (multiple) systems seems to be as high as 60\,\% for the brightest Cepheids \citep{Szabados_2003__0}, and is often neglected in the determination of the P--L relation. The position of a Cepheid in the P--L diagram could be biased by the presence of close bright companions. If the difference in magnitude between the Cepheid and its companion is small, the apparent magnitude of the Cepheid will be overestimated. This has a particular impact on the use of a surface brightness (SB) method to estimate the radius and luminosity \citep[e.g.][]{Fouque_1997_04_0,Gieren_1998_03_0}. The radial velocity measurements can also be altered because of orbital effects. This leads to a bias in the distance estimate, since radial velocities are necessary in the Baade-Wesselink method to evaluate the radius. The knowledge of potential photometric and astrometric biases caused by the companions is therefore important for the distance scale.

\begin{table*}[!ht]
\centering
\caption{Parameters of the Cepheid V1334~Cyg and its close companion.}
\begin{tabular}{ccccccc|ccccccc} 
\hline
\multicolumn{7}{c|}{Primary (Cepheid)}	&	\multicolumn{5}{c}{Secondary\tablefootmark{f}} \\
$\overline{m}_\mathrm{V}$\tablefootmark{a}	&	$\overline{m}_\mathrm{K}$\tablefootmark{b}	&	$\overline{m}_\mathrm{H}$\tablefootmark{b}	&	Sp. Type\tablefootmark{c}	&	$P_\mathrm{pul}$\tablefootmark{c}	&	$\overline{\theta}_\mathrm{LD}$\tablefootmark{d}	&	$d$\tablefootmark{e}	&	Sp. Type	&	$P_\mathrm{orb}$	&	$T_0$ 	&	$e$	&	$a_1\sin{i}$ 	&	$\omega$	&	$f(M)$	\\
	&	&	&	&	(days)	&	(mas)	&	(pc)	&	&	(days)	&	(days)	&	&	(AU)	&	(rad)	&	(M$_\odot$)	\\
\hline
%\noalign{\smallskip}
5.87	&	4.46	&	4.66	&	F2Ib	&	3.333	&	0.534	&	683	&	--	&	1937.5	& 2~443~607	&	0.197	&	2.46	&	3.95	&	0.529	\\
\hline
\end{tabular}
\tablefoot{$\overline{m}_\mathrm{V}$, $\overline{m}_\mathrm{K}$, $\overline{m}_\mathrm{H}$: mean apparent $V$, $K$ and $H$ magnitudes. Sp. Type: spectral type. $P_\mathrm{pul}$: period of pulsation. $\overline{\theta}_\mathrm{LD}$: mean angular diameter. $d$: distance. $P_\mathrm{orb}$: orbital period. $T_0$: time passage through periastron. $e$: eccentricity of the orbit. $a_1\ \sin{i}$: projected semi-major axis of the orbit of the Cepheid about the center of mass of the system. $\omega$: argument of periastron. $f(M)$: spectroscopic mass function.\\
\tablefoottext{a}{from \citet{Klagyivik_2009_09_0}.} 
\tablefoottext{b}{%from \citet{Welch_1984_04_0} for %FF~Aql, 
%U~Aql% and T~Vul
from the 2MASS catalogue \citep{Cutri_2003_03_0}.} 
\tablefoottext{c}{from \citet{Samus_2009_01_0}% for V1334~Cyg, U~Aql and T~Vul. From \citet{Abt_2009_01_0} for FF~Aql
.}
\tablefoottext{d}{%from \citet{Gallenne_2012_05_0} for FF~Aql and T~Vul, and 
from the surface brightness relation of \citet{Kervella_2004_12_0}% for V1334~Cyg and U~Aql
} 
\tablefoottext{e}{%from \citet{Benedict_2007_04_0} for FF~Aql and T~Vul, 
from the P--L relation of %\citet{Storm_2011_10_0} for U~Aql
\citet{Bono_2002_07_0}.} 
\tablefoottext{f}{from \citet{Evans_2000_06_0}. % from \citet{Evans_1990_05_0} and \citet{Benedict_2007_04_0} for FF~Aql,
%and from \citet{Welch_1987_07_0} for U~Aql% and from \citet{Evans_1992_07_0} for T~Vul
.}
%\tablefoottext{g}{true semi-major axis $a$ from \citet{Benedict_2007_04_0} for FF~Aql.}
}
\label{table__system_parameters}
\end{table*}

Cepheids belonging to binary systems also offer the unique opportunity to make progress in resolving the Cepheid mass problem. For many years, stellar evolutionary models have predicted Cepheid masses larger than those derived from pulsation models \citep{Neilson_2011_05_0,Keller_2008_04_0,Bono_2006__0,Bono_1999_05_0}. To investigate the origin of this discrepancy, the combination of spectroscopic and interferometric measurements will allow us to derive orbital elements and dynamical masses. Recent analyses for two eclipsing binary Cepheids in the LMC were carried out by \citet{Pietrzynski_2010_11_0,Pietrzynski_2011_12_0} which yielded masses with an accuracy $< 1.5$\,\%. These dynamical masses are in agreement with the ones calculated from stellar pulsation models, suggesting that the pulsational theory provides the true current masses \citep[see also][]{Prada-Moroni_2012_04_0}. However, this first conclusion needs additional mass estimates to better constrain the two models, particularly for wide binaries for which no significant physical interaction between the stars is expected.

A number of Galactic Cepheids are known to have companions closer than 30\arcsec\ \citep[e.g.][]{Evans_2008_09_0,Remage-Evans_2011_12_0}, but most of them are located too close to the Cepheid ($\sim$1--20\,mas) to be observed with a 10--meter class telescopes. The already existing orbit measurements were estimated only from IUE spectrum or from the radial velocity variations. The only actual way to spatially resolve such systems is to use long-baseline interferometry. We therefore have started a long-term interferometric observing program that aims at studying a sample of seven northern and southern binary Cepheids. The first goal is to determine the angular separation and the apparent brightness ratio from the interferometric visibility and closure phase measurements. Our long-term objective, which needs a good sampling of the orbital period to get a reliable fit (several years), is to determine the full set of orbital elements, absolute masses and geometric distances.

The binary (or multiple) Cepheid systems were selected according to two main criteria. Firstly, the angular separations (mainly estimated from spectroscopy so far) have to be resolvable by the existing long-baseline interferometers. Secondly, the contrast between the Cepheid and its companion should not be too large, to ensure the flux contribution of the companion is detectable in the data. We therefore selected systems with an angular separation $\gtrsim 0.5$\,mas, and a dynamic range $\gtrsim 1:300$.

In this paper, we present the first results for the Cepheid V1334~Cyg, observed with the MIRC combiner in the $H$ band at the CHARA Array. The paper is organized as follows. We first introduce the observed Cepheid, with some knowledge about the primary and its companion. In Sect.~\ref{section__observations_and_data_reduction}, the instrument configurations and the data reduction are detailed. The data analysis, including the interferometric models used and the fitting steps, is discussed in Sect.~\ref{section__model_fitting}. We then combined our interferometric results with spectroscopic data in Sect.~\ref{section__discussion} to derive the parameters of the V1334~Cyg system.

\section{V1334~Cyg}

\object{V1334~Cyg} (HD~203156, HR~8157) was the first Cepheid observed for our program. %, \object{FF~Aql} and \object{T~Vul}. 
We present in Table~\ref{table__system_parameters} the known parameters of this system from the literature. The information about the Cepheid companion were derived from spectroscopic observations \citep{Evans_1995_05_0,Evans_2000_06_0}.

This short-period Cepheid is an interesting system because it has been studied for many years but the binary or triple nature of the system is still debated. It has been suspected to be a member of a visual binary system for decades \citep[see e.g.][]{Millis_1969__0,Abt_1970_04_0}, with a separation between 0.1--0.2\arcsec, but it has not been spatially resolved so far. Early radial velocity measurements also showed strong evidence of a spectroscopic binary, but the orbital period could not be fully constrained. \citet{Abt_1970_04_0} found a period of $\sim 30$\,years for the visual component, but they also noticed a variation in the centre of mass velocity ($v_\gamma$) of a shorter time-scale. The same variation was also observed by \citet{Szabados_1991_01_0}, and is likely linked to a second closer companion. From the International Ultraviolet Exporer (IUE) low-resolution spectra, \citet{Evans_1995_05_0} detected the hottest star in the system, and derived the spectral type to be a B7.0V star. In a subsequent work, \citet{Evans_2000_06_0} concluded that the hottest star is the visual companion. From 30\,years of radial velocity measurements, the same author solved for the orbital parameters of the companion and found a period of 1937\,days (see Table~\ref{table__system_parameters}), that is significantly shorter than the value derived by \citet{Abt_1970_04_0}, strongly suggesting the presence of a third component. They also derived a projected semi-major axis for the orbit of the Cepheid around the center of mass $a_1\,\sin i = 2.46$\,AU (see Table~\ref{table__system_parameters}). \citet{Kiss_2000_05_0} also detected the change in $v_\gamma$, and hypothesized a yellow-bright close companion.

Many authors attempted to resolve the wide component, but we still do not have a firm conclusion because of its intermittent detection. For instance, \citet{Evans_2006_11_0} set an upper limit of $\sim 20$\,mas using the Hubble Space Telescope in the far-ultraviolet (although depending on the brightness of the stars). \citet{Scardia_2008_01_0} mentioned the detection of the companion from speckle interferometry in the $V$ band, and measured a separation of 160\,mas, while most of previous speckle observations failed to resolve the system \citep[from 1976 to 2005, see e.g.][see the last reference for a more detailled discussion]{McAlister_1978_06_0,Hartkopf_1984_01_0,Evans_2006_11_0}, setting a separation $< 35$\,mas.
As argued by \citet{Evans_2006_11_0}, these intermittent detections could be linked to a peculiar orbit, making the visual companion undetectable at some orbital phase, or it could even not exist.

\section{Observations and data reduction}
\label{section__observations_and_data_reduction}

\begin{figure*}[!ht]
\centering
\resizebox{\hsize}{!}{\includegraphics{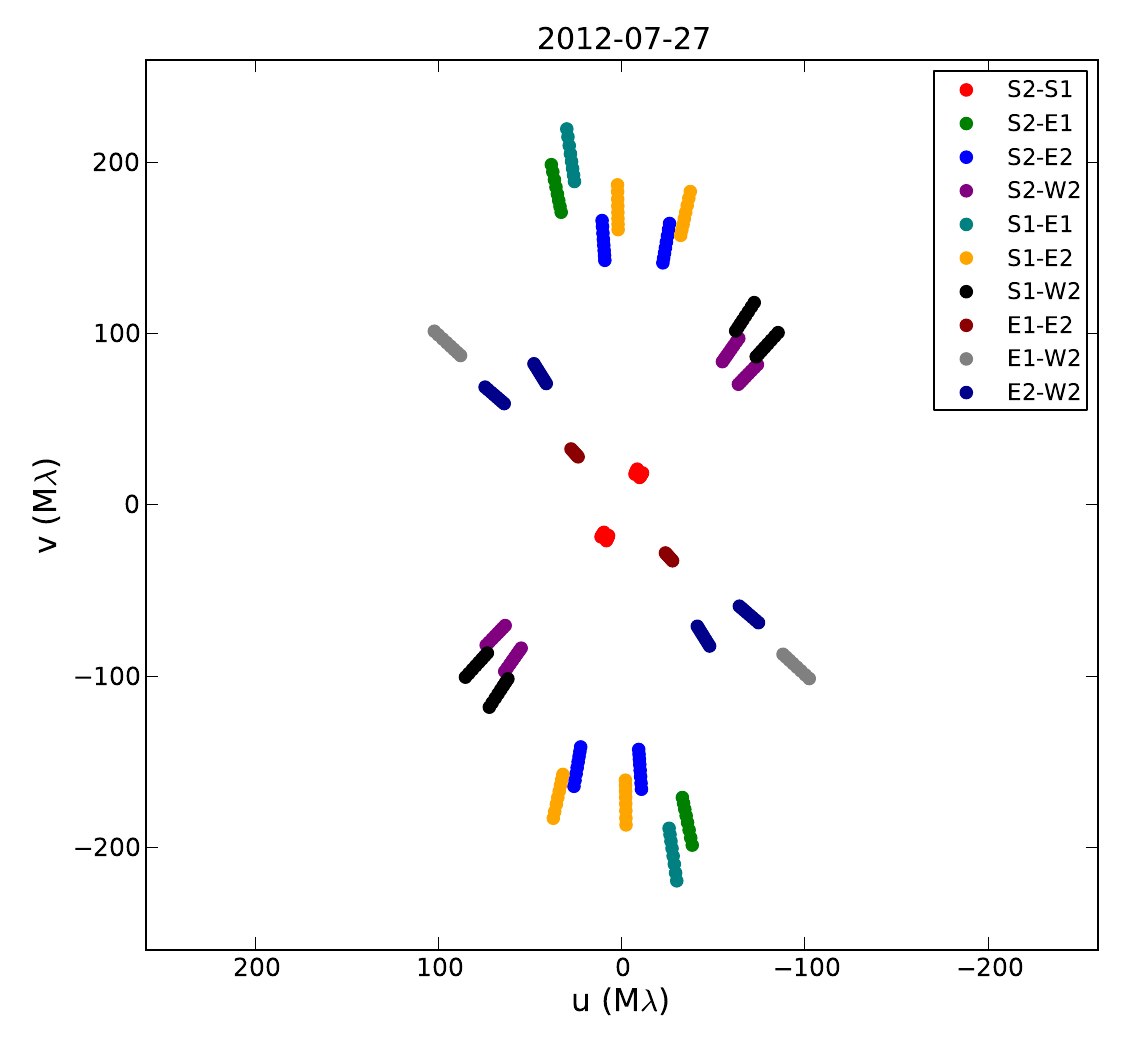}\hspace{.5cm}
\includegraphics{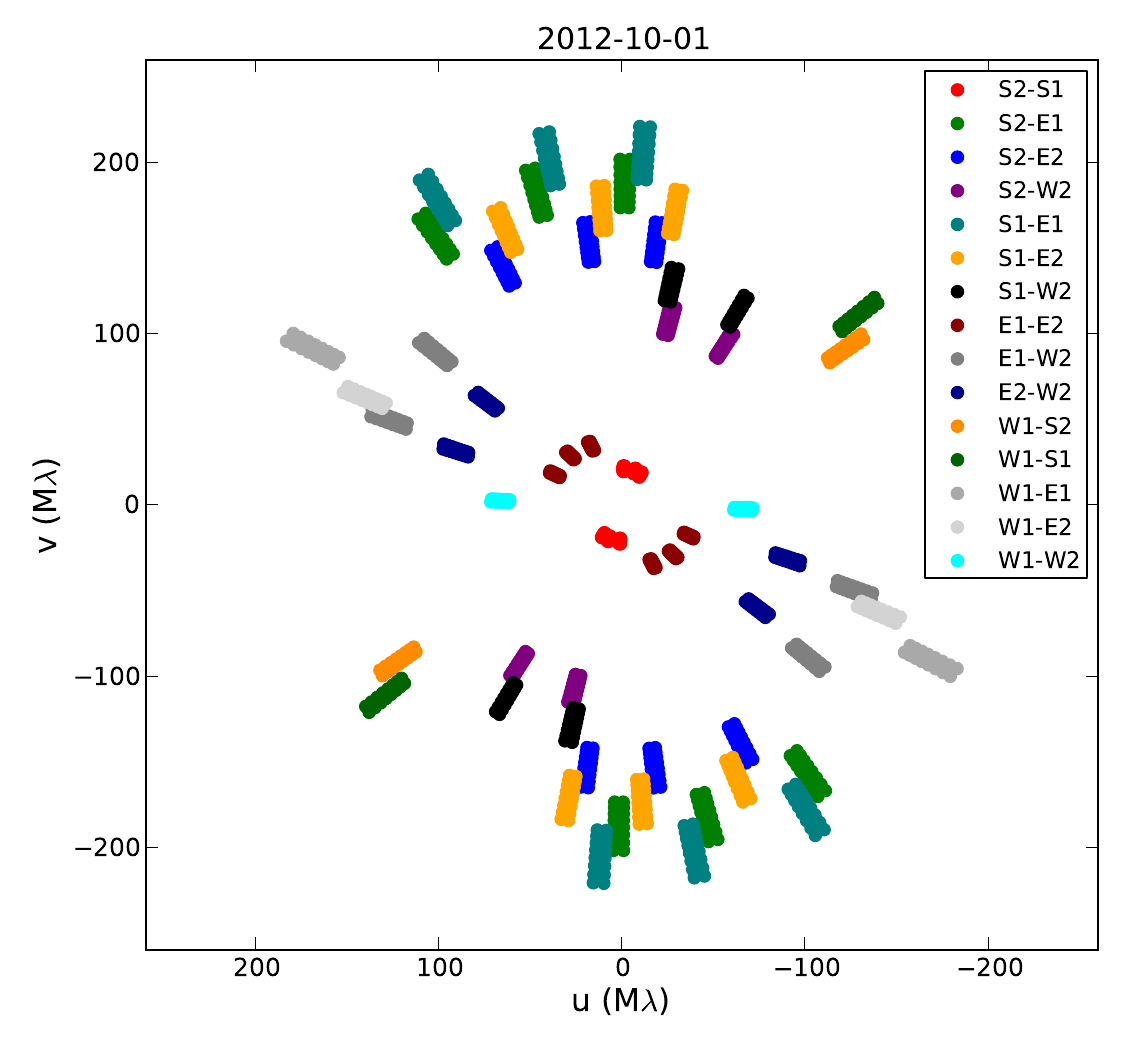}
}
\caption{$(u,v)$ plane coverage of our observations of V1334~Cyg.}
\label{image__uv_plan}
\end{figure*}

Our observations were performed using the Michigan InfraRed Combiner (MIRC), installed at the CHARA array \citep{ten-Brummelaar_2005_07_0} located on Mount Wilson, California. The array consists of six 1\,m aperture telescopes with an Y-shaped configuration (two telescopes on each branch), oriented to the east (E1, E2), west (W1,W2) and south (S1, S2), and so offering a good coverage of the $(u,v)$ plane. The baselines range from 34\,m to 331\,m, providing an angular resolution down to 0.5\,mas in $H$.

MIRC \citep{Monnier_2004_10_0,Monnier_2010_07_0} combines the light coming from all six telescopes in the $K$ or $H$ bands, with three spectral resolutions ($R = 42, 150$ and 400). The recombination of six telescopes gives simultaneously 15 fringe visibilities and 20 closure phase measurements, that are our primary observables.

Our observations were carried out on UT 2012 July 27 and October 1 using an $H$ band filter ($\lambda_0 = 1.65\,\mu$m) and either four, five or six telescopes. We used the low spectral resolution mode, where a prism splits the light on the detector into 8 narrow spectral channels. The $(u,v)$ plane coverage for these nights is presented in Fig.~\ref{image__uv_plan}. We followed a standard observing procedure, i.e we monitored the interferometric transfer function by observing a calibrator before and after our Cepheids. The calibrators were selected using the \textit{SearchCal}\footnote{Available at http://www.jmmc.fr/searchcal.} software \citep{Bonneau_2006_09_0,Bonneau_2011_11_0} provided by the JMMC. The journal of the observations is presented in Table~\ref{table__journal}, and the corresponding calibrators are listed in Table~\ref{table__calibrator}. 

\begin{table}[]
\centering
\caption{Journal of the observations.}
\begin{tabular}{ccc} 
\hline
\hline
UT 																	&	Star				&		Configuration			\\
\hline
%06:40				               &	HD~177756		 &	S1-S2-E2-W1-W2		\\
%07:10 						       &	U~Aql				 &	S1-S2-E2-W1-W2		\\
%07:43							   &	HD~185124		&	S1-S2-E2-W1-W2		\\
%08:07		   					   &	U~Aql				  &	S1-S2-W1-W2		\\
%08:35							   &	HD~198001		&	S1-S2-E2-W1-W2			\\
2012~Jul.~27~09:52							   								&	\object{HD~200577}  		&	S1-S2-E1-E2-W2 		\\ 
\textcolor{white}{2012~Jul.~27}~10:23							   &	V1334~Cyg		 				&	S1-S2-E1-E2-W2 		\\
\textcolor{white}{2012~Jul.~27}~11:08							   &	\object{HD~214200}		&	S1-S2-E1-E2-W2 		\\
\textcolor{white}{2012~Jul.~27}~11:44							   &	V1334~Cyg		 				&	S1-S2-E1-E2 		\\
2012~Oct.~01~02:50															&	\object{HD~185395}		&	S1-S2-E1-E2-W1-W2	\\
\textcolor{white}{2012~Oct.~01}~03:40							& 		V1334~Cyg						&	S1-S2-E1-E2-W2		\\
\textcolor{white}{2012~Oct.~01}~05:01							& 		\object{HD~199956}		&	S1-S2-E1-E2-W1-W2		\\
\textcolor{white}{2012~Oct.~01}~05:44							& 		V1334~Cyg						&	S1-S2-E1-E2-W1-W2		\\
\textcolor{white}{2012~Oct.~01}~06:28							& 		\object{HD~218470}		&	S1-S2-E1-E2-W1-W2		\\
\textcolor{white}{2012~Oct.~01}~07:11							& 		V1334~Cyg						&	S1-S2-E1-E2		\\
\textcolor{white}{2012~Oct.~01}~07:48							& 		\object{HD~207978}		&	S1-S2-E1-E2		\\
%\hline
%2012 Jul 27		 				 &	HD~		 &					\\
\hline
\end{tabular}
\label{table__journal}
\end{table}

\begin{table}[]
\centering
\caption{Calibrators used for our observations.}
\begin{tabular}{cccccc} 
\hline
\hline
Calibrator 													&	$m_\mathrm{V}$	& 	$m_\mathrm{H}$	&	Sp.\;Type	& 	$\theta_\mathrm{UD}$	&	$\gamma$		\\
	(HD)	  															&								&								&						&	(mas)							&	($\degr$)	\\
\hline		
%\multicolumn{6}{c}{V1334~Cyg} \\
200577								& 	6.1	  					&  4.1						&  G8III  				&  $0.758 \pm 0.052$ 		&	3.2			\\ 
214200\tablefootmark{a}		&	6.1						&	4.2						&	K0				&	$0.790 \pm 0.050$		&	15.5			\\
185395\tablefootmark{a}		&	4.5						&	3.7						&	F4V				&	$0.750\pm 0.060$		&	18.2		\\
199956								&	6.6						&	4.5						&	K0				&	$0.603\pm 0.043$		&	7.1		\\
218470									&	5.7						&	4.7						&	F5V				&	$0.477\pm 0.033$		&	15.5		\\
207978									&	5.5						&	4.4						&	F6IV				&	$0.571\pm 0.040$		&	21.6		\\
\hline
\end{tabular}
\tablefoot{$m_\mathrm{V}, m_\mathrm{H}$: magnitudes in $V$ and $H$ bands. $\theta_\mathrm{UD}$: uniform disk angular diameter in $H$ band. $\gamma$: angular distance to the Cepheid.\\
\tablefoottext{a}{Redetermined from internal MIRC calibration with HD~200577.}
}
\label{table__calibrator}
\end{table}

The data were reduced with the standard MIRC pipeline \citep{Monnier_2007_07_0}. The main procedure is to compute squared visibilities and triple products for each baseline and spectral channel, and to correct for photon and readout noises. A recent upgrade includes a simultaneous measurement of the photometric channels with the fringes, enhancing the accuracy of MIRC down to 3\,\% in visibilities \citep{Che_2010_07_0,Che_2012_07_0}. For the July observations, we used a coherent integration of 75\,ms to improve the signal-to-noise ratio of the closure phase, at some expense of the visibility calibration, but for October observations, we used a more standard 17\,ms integration time. We applied the same calibration error model as described in \citet{Monnier_2012_12_0}. We then did an incoherent average of 15\,min for the final data. We note that this average prevent us from detecting a periodic signal from a potential third component, and the detection of any incoherent light would then be limited by the uncertainty in the visibilities.

\section{Model fitting}
\label{section__model_fitting}
To model the squared visibilities,  triple amplitude and closure phase signals, we used the \textit{LITpro}\footnote{LITpro software available at http://www.jmmc.fr/litpro.} model fitting software \citep{Tallon-Bosc_2008_07_0}, based on the Levenberg-Marquardt algorithm. It provides a set of elementary models that can be combined all together. The software also contains a tool allowing the search for the global minimum to solve for the problem of multiple $\chi^2$ minima.

The two epoch observations (July and October) were reduced separately in order to detect the changing position of the companion.

\subsection{The models}

As a first step, the primary component (the Cepheid) was modeled with a uniform disk (UD) angular diameter. The complex visibility model is:
\begin{displaymath}
V_\star(u,v) = \frac{2J_\mathrm{1}(x)}{x},
\end{displaymath}
with $J_\mathrm{1}(x)$ the first-order Bessel function, $x = \pi \theta_\mathrm{UD}\sqrt{u^2 + v^2}/\lambda$, $(u,v)$ the spatial frequencies, $\theta_\mathrm{UD}$ the UD angular diameter, and $\lambda$ the wavelength.

The choice of a UD diameter instead of a limb-darkened (LD) disk for the fitting procedure is justified because the angular diameters of the Cepheids are small compared to the angular resolution of the interferometer, and the limb darkening effects are therefore undetectable. The conversion from UD to LD angular diameter was done by using a linear-law parametrization $I_\lambda (\mu) = 1 - u_\lambda(1 - \mu)$, with the LD coefficient $u_\lambda$  from \citet{Claret_2011_05_0}. The conversion is then given by the approximation \citep{Hanbury-Brown_1974_06_0}:
\begin{displaymath}
\theta_\mathrm{LD}(\lambda) = \theta_\mathrm{UD}(\lambda) \sqrt{\frac{1-u_\lambda/3}{1-7u_\lambda/15}}
\end{displaymath}

It is worth mentioning that the uncertainty of the limb-darkened coefficient has a small impact on the angular diameter conversion. A variation of 20\,\% of $u_\lambda$ gives a LD diameter difference of less than 0.5\,\%.

As a second step, we fitted to the data a model with one companion, assumed to be unresolved by the array. The corresponding model for this binary system is:
\begin{displaymath}
V(u,v) = \frac{V_\star(u,v) + f\,V(u,v)}{1 + f},
\end{displaymath}
where $f$ is the flux ratio between the companion and the Cepheid, and $V(u,v)$ is the complex visibility model of an unresolved source:
\begin{displaymath}
V(u,v) = \exp(-2i\pi (u\Delta \alpha + v\Delta \delta)/\lambda),
\end{displaymath}
with $(\Delta \alpha,\Delta \delta)$ the relative position of the companion w.r.t the Cepheid.

The closure phase is then estimated from the modulus and argument of the bispectrum, $B_{123}$, for each closed baseline triangle and spectral channel:
\begin{align*}
& | B_{123} | = | V(u_\mathrm{1},v_\mathrm{1})V(u_\mathrm{2},v_\mathrm{2})V^\ast(u_\mathrm{3},v_\mathrm{3}) |\\
& \arg(B_{123}) = \phi_\mathrm{123} = \arg(V(u_\mathrm{1},v_\mathrm{1})V(u_\mathrm{2},v_\mathrm{2})V^\ast(u_\mathrm{3},v_\mathrm{3}))
\end{align*}

We assumed that the variation of the angular diameter between different acquisitions in the same night is negligible compared to our level of accuracy.

\subsection{Fitting steps}
\label{subsection__fitting_steps}

We performed a least-squares model fit simultaneously with the squared visibility, triple amplitude and closure phase measurements. Our search strategy was the following. We first proceeded to a grid search in the $\chi^2$ space to determine the approximate position and brightness ratio of the companion. A first grid search between $\pm 20$\,mas with a 0.2\,mas spacing was performed with various flux ratio (from $0 < f < 0.2$ with 0.005 steps). Then a second grid of $\pm 1$\,mas with a 0.01\,mas spacing around the most likely position and brightness ratio was used to refine the position. Finally all parameters were fitted using the refined values. The first guess for the angular diameter was taken from Table~\ref{table__system_parameters}.

\begin{table}[]
\centering
\caption{Summary of the parameters estimated from the model fit.}
\begin{tabular}{ccccc} 
\hline
\hline
	  															& 	2012-07-27					&  2012-10-01			\\
\hline
Single star model \\
$\theta_\mathrm{UD}$ (mas)			 &  $0.565 \pm 0.052$		&	$0.487 \pm 0.045$       	\\
$\theta_\mathrm{LD}$ (mas)			 &  $0.575 \pm 0.052$		&  $0.496 \pm 0.045$		          	\\
$\chi^2_r$									   &  1.63		          				&	2.08            								\\
\hline
Binary model \\
$\theta_\mathrm{UD}$ (mas)				 &  $0.494 \pm 0.053$		&	$0.436 \pm 0.045$        	\\
$\theta_\mathrm{LD}$ (mas)				 &  $0.503 \pm 0.053$		&	$0.444 \pm 0.045$	         	\\
$f$	(\%)												& $3.15 \pm 0.15$ 			&	$3.08 \pm 0.09$      		\\
$\Delta \alpha$ (mas)						  &  $-1.153  \pm 0.030$	&	$-0.113 \pm 0.014$   	 	\\
$\Delta \delta$ (mas)				  		&  $-8.836  \pm 0.017$	&	$-8.359  \pm 0.009$	   	\\
$\chi^2_r$								          &  0.34		                          &	1.24  									\\
\hline
\end{tabular}
\tablefoot{$\theta_\mathrm{UD}$ and $\theta_\mathrm{LD}$ are the uniform and limb-darkened disk angular diameter, respectively. $f$, $x$ and $y$ correspond to the flux ratio and position of the companion. $\chi^2_r$ is the reduced $\chi^2$ of the corresponding fitted model.}
\label{table__fitted_parameters}
\end{table}

\subsection{Results}

\begin{figure*}[!ht]
\centering
\resizebox{\hsize}{!}{\includegraphics{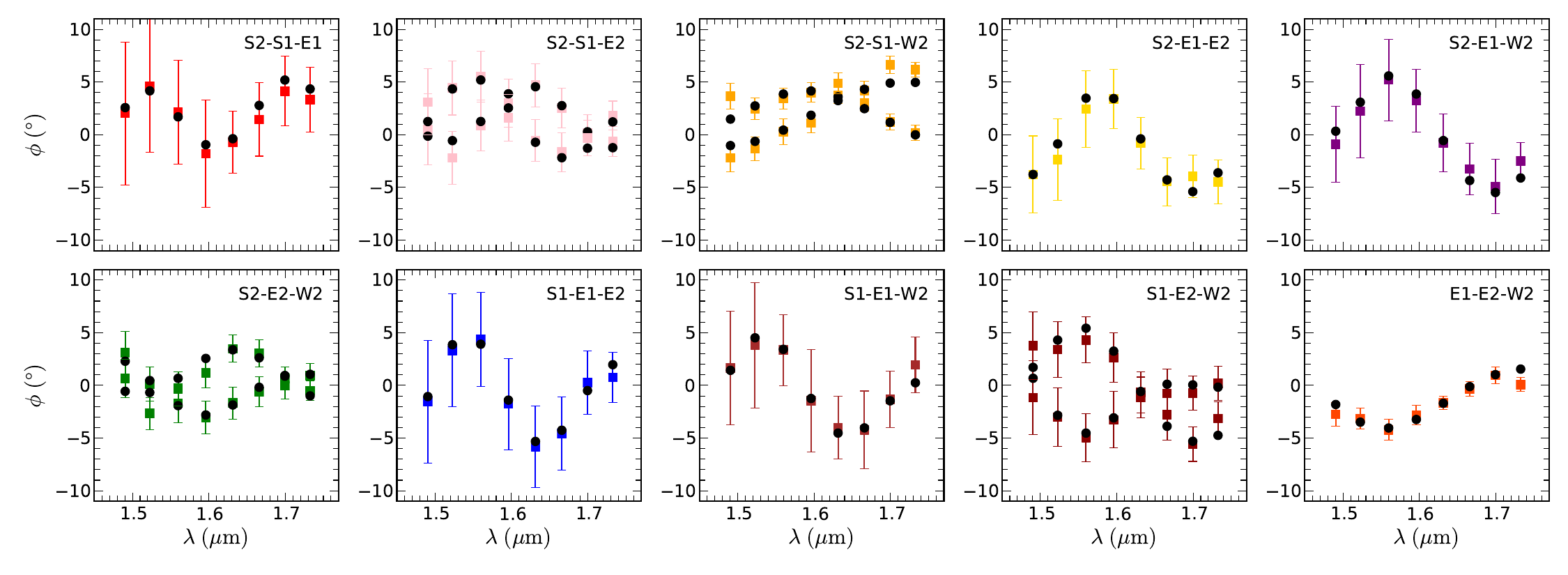}}
\caption{Closure phase signal for the first epoch. The color-coded squares are the data, while the black dots represent the binary model.}
\label{image__cp_v1334cyg_1}
\end{figure*}
\begin{figure*}[!ht]
\centering
\resizebox{\hsize}{!}{\includegraphics{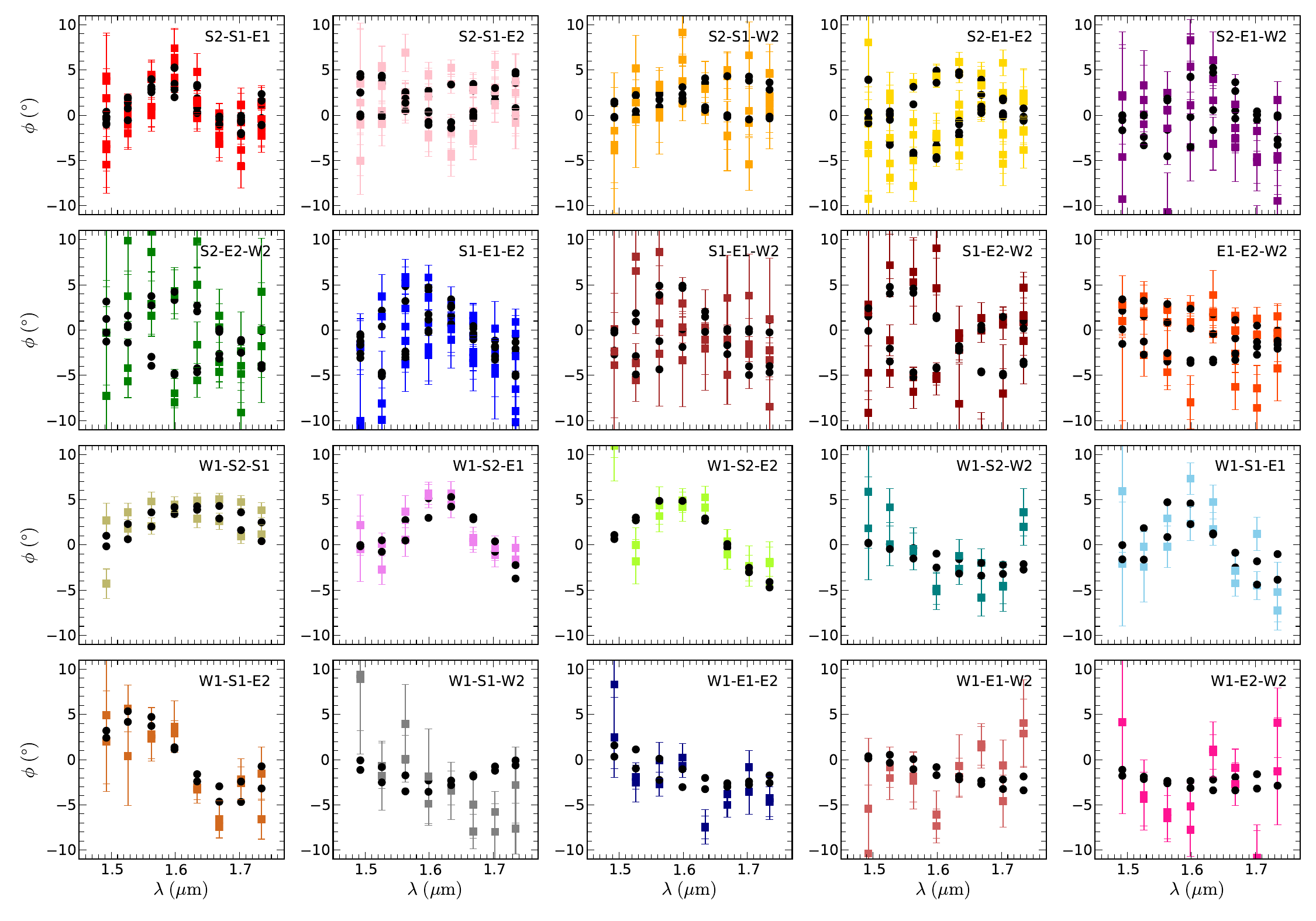}}
\caption{Closure phase signal for the second epoch. The color-coded squares are the data, while the black dots represent the binary model.}
\label{image__cp_v1334cyg_2}
\end{figure*}

\begin{figure*}[]
\centering
\resizebox{\hsize}{!}{\includegraphics{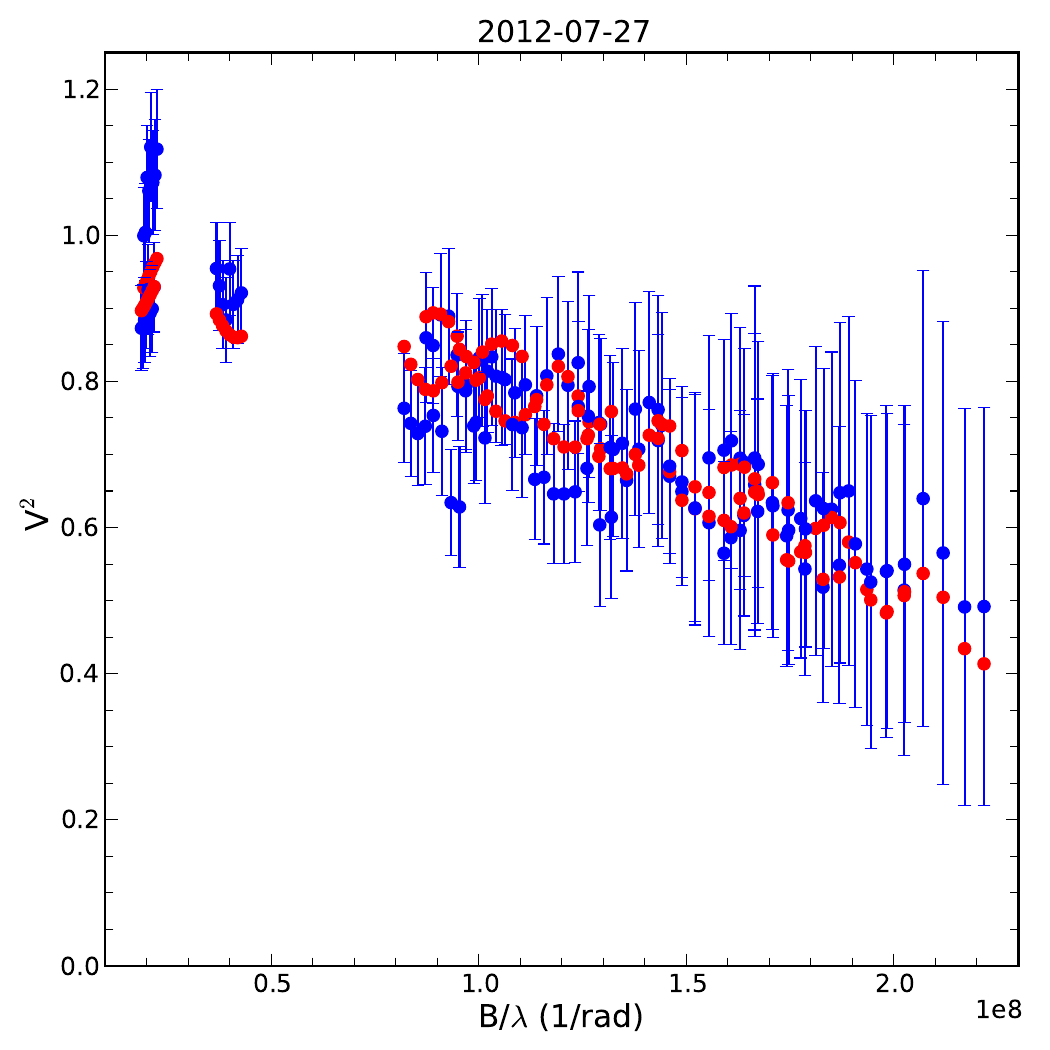}
\includegraphics{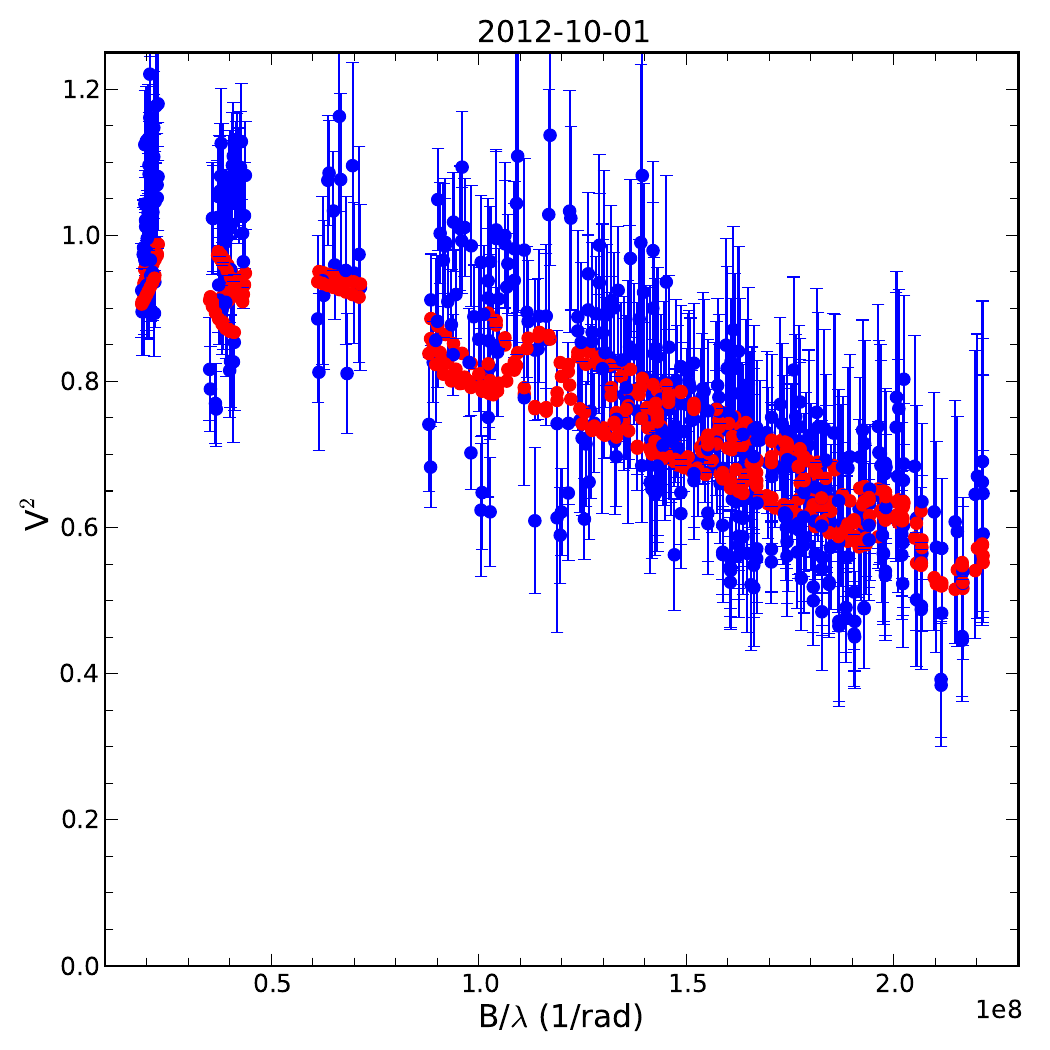}
}
\caption{Squared visibility measurements for the two epochs. The data are represented in blue, while the red dots are the fitted binary model.}
\label{image__visibility}
\end{figure*}

The closure phase signal and squared visibilities, presented in Figs.~\ref{image__cp_v1334cyg_1}, \ref{image__cp_v1334cyg_2} and \ref{image__visibility}, clearly show a departure from a single star with a symmetric brightness distribution. These variations reveal the presence of at least one companion.

\begin{figure*}[]
\centering
\resizebox{\hsize}{!}{\includegraphics{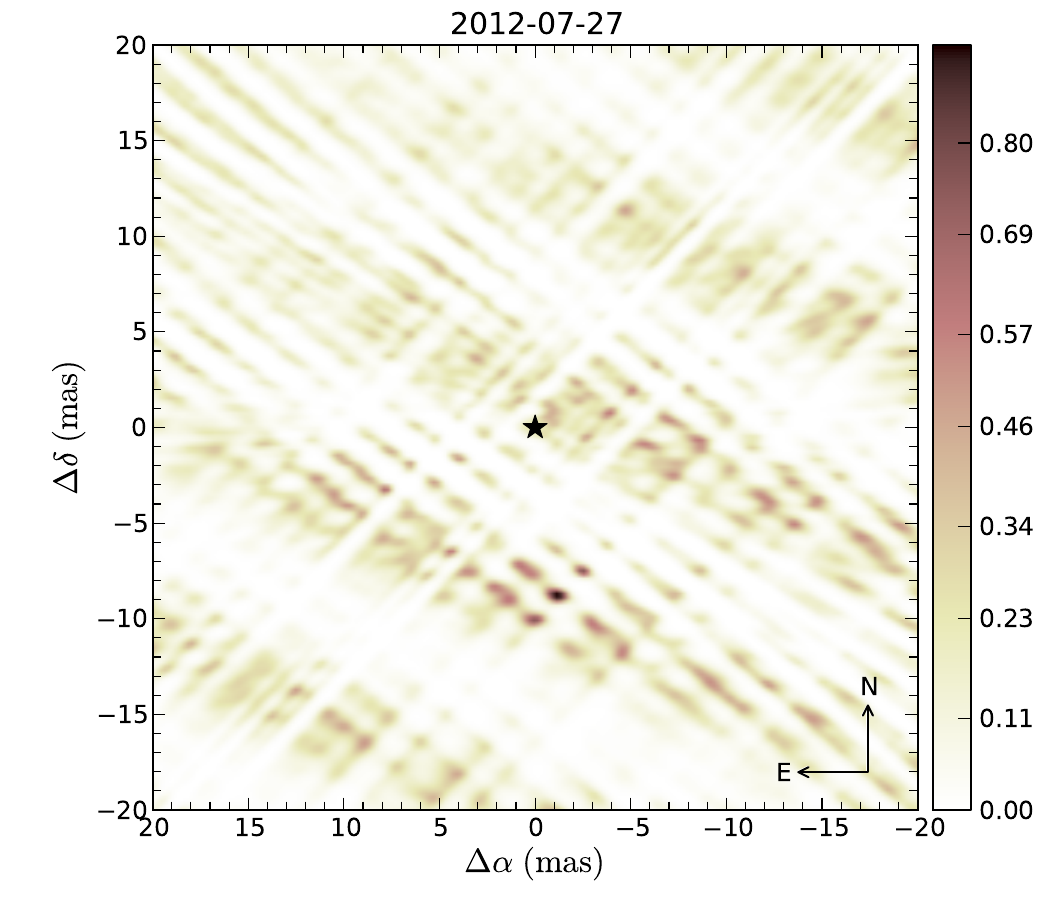}
\includegraphics{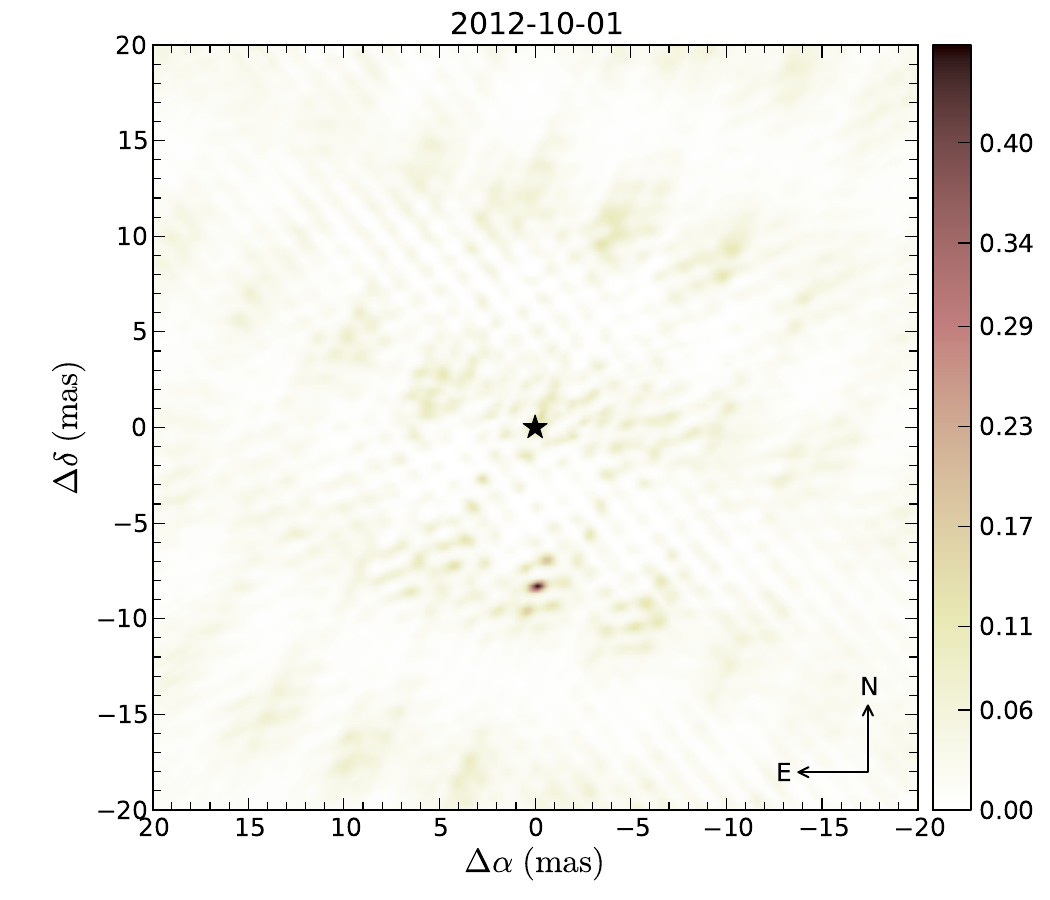}}
\caption{Probability maps for the two observed epochs, showing the maximum obtained at each point of the search region with a 3\,\% flux ratio.}
\label{image__chi2_map}
\end{figure*}

We chose as first guess $\theta_\mathrm{UD} = 0.547$\,mas, and performed the fitting procedure with a grid search as explained in Sect.~\ref{subsection__fitting_steps}. The probability maps for our two epochs of observations are shown in Fig.~\ref{image__chi2_map}, and the fitted parameters are reported in Table~\ref{table__fitted_parameters}. The model is also represented graphically with black dots in Figs.~\ref{image__cp_v1334cyg_1}, \ref{image__cp_v1334cyg_2}  and red dots in Fig.~\ref{image__visibility}. We notice a good agreement between the model and the data. The fitted model give the most probable location of this companion at an angular separation $\rho = 8.91$\,mas and a position angle $PA = -172.6\degr$ for the first epoch, and $\rho = 8.36$\,mas and $PA = -179.2\degr$ for the second. The measured flux ratio is also particularly consistent between the two epochs.

We estimated the limb-darkened angular diameter to be $\theta_\mathrm{LD} = 0.503 \pm 0.053$\,mas for the first epoch, using the limb-darkening coefficient $u_\lambda = 0.2423$ \citep{Claret_2011_05_0}, chosen from the stellar parameters: $T_\mathrm{eff} = 6250$\,K, $\log g = 2.0$, [Fe/H] = 0.0 and $v_\mathrm{t} = 4$\,m~s$^\mathrm{-1}$ \citep[][at pulsation phase $\phi = 0.17$]{Luck_2008_07_0}. For the second epoch, the photospheric parameters did not change significantly (because it is low-amplitude pulsation Cepheid), leading to a same $u_\lambda$ to find $\theta_\mathrm{LD} = 0.444 \pm 0.045$\,mas (at pulsation phase $\phi = 0.91$). These measurements are in agreement with the averaged angular diameters predicted from a surface brightness relation \citep[$\sim 0.545$\,mas,][]{Kervella_2004_10_0,Moskalik_2005_06_0}, taking into account the small amplitude variation. As no IR photometric curves are available, it was not possible to compare our value with an angular diameter derived from a SB technique for the actual phase of our measurements.

\section{Combination of spectroscopy and interferometry}
\label{section__discussion}

\subsection{Preliminary orbit}

\begin{figure*}[!ht]
\centering
\resizebox{\hsize}{!}{\includegraphics{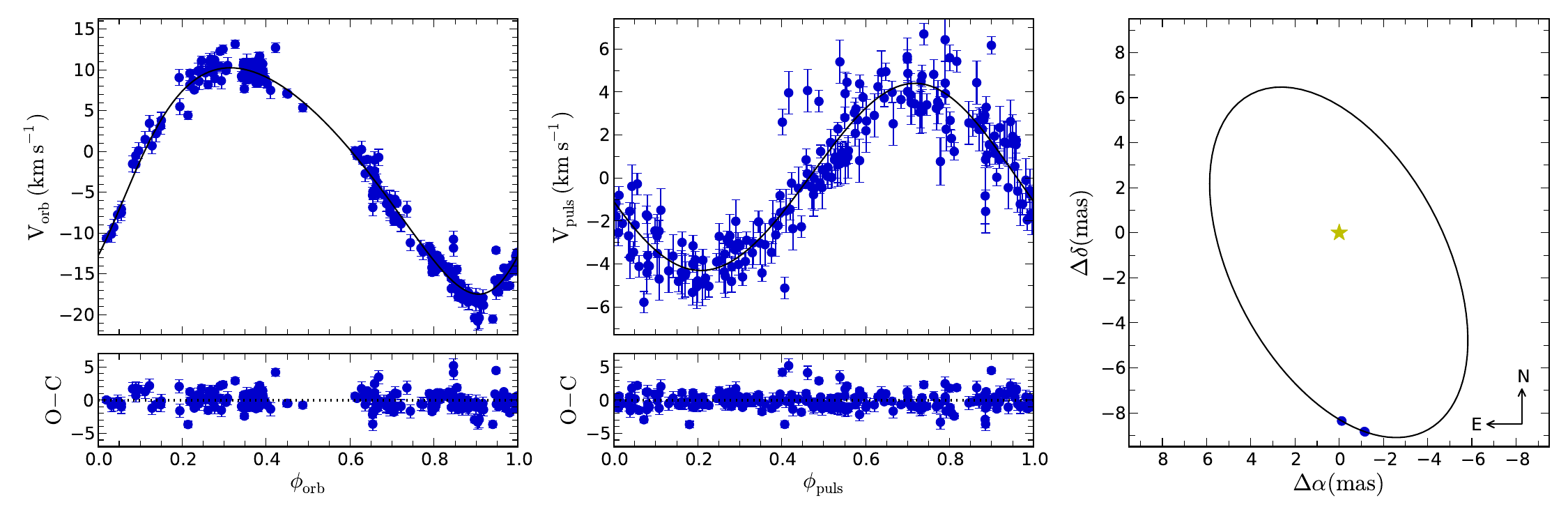}}
\caption{\textit{Left}: fitted (solid line) and measured (blue dots) orbital velocity. \textit{Middle}: fitted (solid line) and measured (blue dots) pulsation velocity. \textit{Right}: orbit of V1334~Cyg Ab. The data points are the MIRC results from Table~\ref{table__fitted_parameters}.}
\label{image__rv}
\end{figure*}

Because of its intermittent detections, the existence of the wide component is rather uncertain, and might not exist. We claim that the companion detected with MIRC and IUE \citep{Evans_1995_05_0} can be identified as the spectroscopic one. Therefore, our measured separations should be compatible with the spectroscopic orbit, and it should be possible to derive all the orbital parameters.

We therefore combined our astrometric measurements with the single-line radial velocity data gathered by \citet{Evans_2000_06_0}, in order to derive the complete orbital elements for V1334~Cyg Ab. We stress that this analysis is preliminary as we have only two astrometric measurements so far. We computed the orbital solutions through the formalism developed by \citet{Wright_2009_05_0}, slightly modified to include the pulsation velocity of the form:
\begin{displaymath}
V_\mathrm{puls}(t) = \sum_\mathrm{i=1}^\mathrm{2} A_\mathrm{i}\cos (2i\pi \phi(t) + B_\mathrm{i})
\end{displaymath}
where $A_\mathrm{i}$ and $B_\mathrm{i}$ are the fitted Fourier coefficients, and $\phi$ denotes the pulsation phase. 

We solved for all the orbital elements ($P_\mathrm{orb}, T_\mathrm{p}, e, K, v_\gamma, \omega, a, i, \Omega$) and pulsation parameters ($P_\mathrm{puls}, A_1, A_2, B_1, B_2$). The reference epoch of maximum light for the Cepheid, $T_0$, was held fixed to the value given by \citet{Samus_2009_01_0}. The initial values for the orbit and the pulsation were chosen from \citet{Evans_2000_06_0} and \citet{Samus_2009_01_0}. The final elements derived from our combined fit are listed in Table~\ref{table__fitted_orbit}. The fitted parameters are in good agreement with those from \citet{Evans_2000_06_0}. The quoted uncertainties for the elements derived from spectroscopy ($P_\mathrm{orb}, T_\mathrm{p}, e, K, v_\gamma, \omega,$) were estimated using the bootstrapping technique (with replacement and 500 bootstrap samples). For the remaining elements derived from interferometry ($a, i$ and $\Omega$), we refitted the orbits 500 times, each time adding Gaussian noise to each astrometric point according to their uncertainties. The standard deviation from these trials is then used as the uncertainty. Fig.~\ref{image__rv} (left and middle) shows the orbital and pulsation velocities disentangled from the radial velocity measurements. The solid black lines denote our fitted curves. The final best fit orbit of V1334~Cyg Ab is also plotted in Fig.~\ref{image__rv} (right) with our MIRC measurements marked by the blue dot symbols.

\begin{table}[]
\centering
\caption{Fitted orbital and pulsation parameters of V1334~Cyg Ab.}
\begin{tabular}{ccc} 
\hline
\hline
																		& 	Spectroscopy only				&	This work	\\
																		 &	\citep{Evans_2000_06_0}	   &						\\
\hline
Orbit																&												 &						\\	
$P_\mathrm{orb}$ (days)									& $1937.5 \pm  2.1$ 			    & $1938.6 \pm 1.2$ 		\\
$T_\mathrm{p}$ (HJD)									&	$2~443~607 \pm 14$	& $2~443~616.1 \pm 7.3$ \\
$e$																     &	$0.197 \pm 0.009$			   &  $0.190 \pm 0.013$	\\
$K_1$ ($\mathrm{km~s^{-1}}$)					  &	$14.1 \pm 0.1$					 &	$13.86 \pm 0.17$		\\
$v_\gamma$	($\mathrm{km~s^{-1}}$)			&	$-1.8 \pm 0.1$			 	   &	$-1.9 \pm 0.1$			\\
$\omega$	($\degr$)									 &	$226.3 \pm 2.9$				  &	$228.7 \pm 1.6$		\\
$\Omega$	($\degr$)									 &	--										 &	$206.3 \pm 9.4$		\\
$a$ (mas)														 &	--										 &	$8.54 \pm 0.51$		\\
$i$ ($\degr$)													&	--										&	$124.7 \pm 1.8$		\\
$m_\mathrm{H}$											   &	--									  &		$8.47 \pm 0.15$		\\
\hline
Pulsation		& &	\\
$P_\mathrm{puls}$ (days)							  &	$3.33251 \pm 0.00001$ 	&  $3.33250 \pm 0.00002$\\
$T_0$\tablefootmark{a} (HJD)						   & $2~440~124.5330$		&  $2~440~124.5330$		\\
$A_1$															  & --								   &	$4.35 \pm 0.15$			\\
$A_2$															  & --								   &	$1.81 \pm 0.11$		\\
$B_1$															   & --									&	$0.08 \pm 0.06$		\\
$B_2$															   & --									&	$2.72 \pm 1.30$		\\
\hline& 
\end{tabular}
\tablefoot{$P_\mathrm{orb}$: orbital period. $T_\mathrm{p}$: time passage through periastron. $e$: eccentricity. $K$: radial velocity semi-amplitude of the primary. $v_\gamma$: systemic velocity. $\omega$: argument of periastron. $\Omega$: position angle of the ascending node. $a$: semi-major axis. $i$: orbital inclination. $m_\mathrm{H}$: apparent magnitude in $H$. $P_\mathrm{puls}$: pulsation period. $T_0$: reference epoch of maximum light. $A_\mathrm{i}, B_\mathrm{i}$: Fourier parameters.\\
\tablefoottext{a}{From \citet{Samus_2009_01_0}, and held fixed when fitting}
}
\label{table__fitted_orbit}
\end{table}

\subsection{Apparent magnitude, spectral type and mass of the companion}

%A companion (the brightest) is clearly detected in the closure phase signal, and confirms, at least, the binary nature of the V1334~Cyg system. 
Combining the $H$-band magnitude $m_\mathrm{H} = 4.66 \pm 0.04$ given by the 2MASS catalog \citep{Cutri_2003_03_0} with our averaged measured flux ratio $f = 3.10 \pm 0.08$\,\%, we derived for the companion a magnitude $m_\mathrm{H}(\mathrm{comp}) = 8.47 \pm 0.15$, and $m_\mathrm{H}(\mathrm{cep}) = 4.70 \pm 0.15$\,mag. As no $H$-band light curve is available to estimate the Cepheid magnitude at our pulsation phase, an additional uncertainty of 3\,\% was quadratically added to take into account the phase mismatch. The choice of these 3\,\% is based on the amplitude variation of the light curve in $V$ \citep[][that is surely to be lower in $H$]{Klagyivik_2009_09_0}. 

\begin{figure}[!t]
\centering
\resizebox{\hsize}{!}{\includegraphics{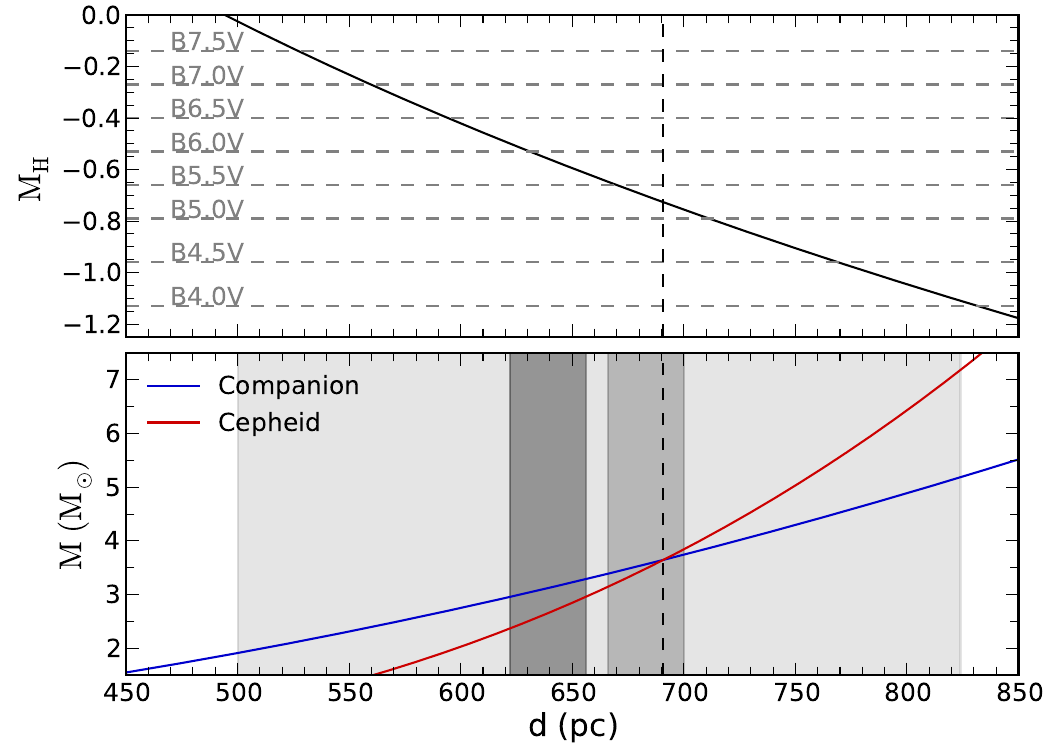}
}
\caption{Absolute magnitude, spectral type of the companion (top panel), and mass of each component (bottom panel) as function of the distance. The shaded gray areas denote the distance range of Hipparcos, F mode P--L relation (left), and FO mode P--L relation (right). The vertical dotted line represents the minimum distance where $M_1 = M_2$.}
\label{image__sp_type}
\end{figure}

The absolute magnitude, $M_\mathrm{H}$, can be estimated knowing the distance to the system. However, there is no accurate determination of the distance for this Cepheid. The Hipparcos data give a distance $d = 662 \pm 162$\,pc \citep{van-Leeuwen_2007_11_0}. The use of a $K$-band P--L relation for first overtone (FO) pulsators \citep[][non-canonical model]{Bono_2002_07_0} gives $d = 683 \pm 17$\,pc, while converting the overtone period to the fundamental one with the period ratio data from \citet{Alcock_1995_04_0} and using a $K$-band P--L relation for fundamental (F) mode pulsator \citep{Storm_2011_10_0} gives $d = 639 \pm 17$\,pc.  As there is no optimum value, we plotted in Fig.~\ref{image__sp_type} (top panel) the spectral type vs. the distance. The previous cited distance ranges set a spectral type for the companion between a B8.0V and B4.0V star. The extinction was assumed negligible at our observing wavelength ($A_\mathrm{H} = 0.023$\,mag, estimated using the total-to-selective absorption ratios $R_\mathrm{V} = 3.1$, $R_\mathrm{H} = A_\mathrm{H}/E(B - V) = R_\mathrm{V}/6.82$ from \citealt{Fouque_2003__0}, and the average color excess $E(B - V) = 0.05$ from \citealt{Evans_1995_05_0} and \citealt{Kovtyukh_2008_09_0})

%f = 3.10+/-0.07

We can also estimate the mass ratio, $q$, for a range of distances by combining the Kepler's third law with the spectroscopic mass function,
\begin{align}
& M_\mathrm{T} = M_1 + M_2 = \dfrac{a^3d^3}{P_\mathrm{orb}^2}, \label{eq__1}\\
& \dfrac{(M_2\sin i)^3}{(M_1 + M_2)^2} = 3.784\times10^{-5}~K_1^3 P_\mathrm{orb} (1 - e^2)^{3/2},\label{eq__2}
\end{align}
which yields:
\begin{displaymath}
q = \dfrac{M_2}{M_1} = \left[ \dfrac{a\,d \sin i}{0.03357 K_1P_\mathrm{orb} \sqrt{1 - e^2}} - 1 \right]^{-1},
\end{displaymath}
with $a$ in arcseconds, $d$ in pc, $P_\mathrm{orb}$ in yr, and $K_1$ in $\mathrm{km\,s^{-1}}$. Combining Eq.~\ref{eq__1} and \ref{eq__2} gives the mass of each component:
\begin{align}
& M_2 = 0.03357 \dfrac{K_1 a^2 d^2 \sqrt{1 - e^2}}{P_\mathrm{orb} \sin i},\label{eq__3}\\
& M_1 = M_\mathrm{T} - M_2. \label{eq__4}
\end{align}

We plotted Eq.~\ref{eq__3} and \ref{eq__4} vs. the distance in Fig.~\ref{image__sp_type} (bottom panel). The shaded gray areas denote the distance range of Hipparcos, F mode P--L relation, and FO mode P--L relation. As we expect a lower mass companion \citep{Evans_1995_05_0}, we can set a lower limit to the distance, $q \leqslant 1$ implies $d \gtrsim 691$\,pc, plotted as a dotted vertical line in Fig.~\ref{image__sp_type}. This limit is consistent with the distance derived from the FO P--L relation \citep[$d = 683$\,pc,][]{Bono_2002_07_0}, and would mean a mass ratio equal to 1 for this system. With this limit, Eq.~\ref{eq__3} yields a minimum mass for the companion (and so the Cepheid) $M_2 \backsimeq 3.6\,M_\odot$, and  a spectral type earlier than a B5.5V star.

\section{Conclusion}

We presented new multiple telescope interferometric observations of the classical Cepheid V1334~Cyg in the $H$ band. For the first time, we were able to spatially resolve the companion at two epochs. We derived the limb-darkened angular diameter for the Cepheid at their corresponding pulsation phase, the relative positions, and the flux ratio of the companion. We combined our accurate astrometric measurements with existing spectroscopic data and derived preliminary orbital solutions for the V1334~Cyg system. We also determined a minimal distance to the system to be $d \sim 691$\,pc, which yields to a minimum mass for the component $M_2 \sim 3.6\,M_\odot$. We also found that its spectral type is earlier than a B5.5V star.

Our work, using multi-telescope recombination, provided unique and useful informations, both on the Cepheid and its companion. These innovative results show the capabilities of long-baseline interferometry to study Cepheids in binary or multiple systems. This is particularly important as most of the companions are located too close to the star to be spatially resolved by a single telescope. This technique provides new observables that can be efficiently combined with spectroscopic results to provide innovative constraints on the system properties. We plan to continue our observing program in the next years to increase the sample of binary Cepheids with well determined orbital elements.

%--------------------ACKNOWLEDGEMENTS--------------------

\begin{acknowledgements}
The authors would like to thank the CHARA Array and Mount Wilson Observatory staff for their support. Research conducted at the CHARA Array is funded by the National Science Foundation through NSF grant AST-0908253, by Georgia State University, the W. M. Keck Foundation, the Packard Foundation, and the NASA Exoplanet Science Institute. JDM acknowledges funding from the NSF grants AST-0707927 and AST-0807577. WG and GP gratefully acknowledge financial support for this work from the BASAL Centro de Astrof\'isica y Tecnolog\'ias Afines (CATA) PFB-06/2007. Support from the Polish National Science Centre grant MAESTRO DEC-2012/06/A/ST9/00269 and the Polish Ministry of Science grant Ideas Plus (awarded to G.~P.) is also acknowledge. This research received the support of PHASE, the high angular resolution partnership between ONERA, Observatoire de Paris, CNRS, and University Denis Diderot Paris 7. A.~G. acknowledges support from FONDECYT grant 3130361. LSz has been supported by the ESTEC Contract No.4000106398/12/NL/KML. This work made use of the SIMBAD and VIZIER astrophysical database from CDS, Strasbourg, France and the bibliographic informations from the NASA Astrophysics Data System. This research has made use of the Jean-Marie Mariotti Center \texttt{LITpro} service, co-developed by CRAL, LAOG and FIZEAU, and \texttt{SearchCal} service, co-developed by FIZEAU and LAOG/IPAG, and of CDS Astronomical Databases SIMBAD and VIZIER.
\end{acknowledgements}

%--------------------BIBLIOGRAPHY--------------------

\bibliographystyle{aa}   % if natbib is available
\bibliography{/Users/alex/Sciences/Articles/bibliographie}

\begin{thebibliography}{51}
\expandafter\ifx\csname natexlab\endcsname\relax\def\natexlab#1{#1}\fi

\bibitem[{{Abt} \& {Levy}(1970)}]{Abt_1970_04_0}
{Abt}, H.~A. \& {Levy}, S.~G. 1970, \pasp, 82, 334

\bibitem[{{Alcock} {et~al.}(1995){Alcock}, {Allsman}, {Axelrod}, {Bennett},
  {Cook}, {Freeman}, {Griest}, {Marshall}, {Peterson}, {Pratt}, {Quinn},
  {Reimann}, {Rodgers}, {Stubbs}, {Sutherland}, \& {Welch}}]{Alcock_1995_04_0}
{Alcock}, C., {Allsman}, R.~A., {Axelrod}, T.~S., {et~al.} 1995, \aj, 109, 1653

\bibitem[{{Barnes}(2009)}]{Barnes_2009_09_0}
{Barnes}, T.~G. 2009, in AIPC Series, ed. {J.~A.~Guzik \& P.~A.~Bradley}, Vol.
  1170, 3--12

\bibitem[{{Bonneau} {et~al.}(2006){Bonneau}, {Clausse}, {Delfosse}, {Mourard},
  {Cetre}, {Chelli}, {Cruzal{\`e}bes}, {Duvert}, \& {Zins}}]{Bonneau_2006_09_0}
{Bonneau}, D., {Clausse}, J.-M., {Delfosse}, X., {et~al.} 2006, \aap, 456, 789

\bibitem[{{Bonneau} {et~al.}(2011){Bonneau}, {Delfosse}, {Mourard}, {Lafrasse},
  {Mella}, {Cetre}, {Clausse}, \& {Zins}}]{Bonneau_2011_11_0}
{Bonneau}, D., {Delfosse}, X., {Mourard}, D., {et~al.} 2011, \aap, 535, A53

\bibitem[{{Bono} {et~al.}(2006){Bono}, {Caputo}, \&
  {Castellani}}]{Bono_2006__0}
{Bono}, G., {Caputo}, F., \& {Castellani}, V. 2006, \memsai, 77, 207

\bibitem[{{Bono} {et~al.}(2010){Bono}, {Caputo}, {Marconi}, \&
  {Musella}}]{Bono_2010_05_0}
{Bono}, G., {Caputo}, F., {Marconi}, M., \& {Musella}, I. 2010, \apj, 715, 277

\bibitem[{{Bono} {et~al.}(2002){Bono}, {Groenewegen}, {Marconi}, \&
  {Caputo}}]{Bono_2002_07_0}
{Bono}, G., {Groenewegen}, M.~A.~T., {Marconi}, M., \& {Caputo}, F. 2002,
  \apjl, 574, L33

\bibitem[{{Bono} {et~al.}(1999){Bono}, {Marconi}, \&
  {Stellingwerf}}]{Bono_1999_05_0}
{Bono}, G., {Marconi}, M., \& {Stellingwerf}, R.~F. 1999, \apjs, 122, 167

\bibitem[{{Caputo} {et~al.}(2005){Caputo}, {Bono}, {Fiorentino}, {Marconi}, \&
  {Musella}}]{Caputo_2005_08_0}
{Caputo}, F., {Bono}, G., {Fiorentino}, G., {Marconi}, M., \& {Musella}, I.
  2005, \apj, 629, 1021

\bibitem[{{Che} {et~al.}(2012){Che}, {Monnier}, {Kraus}, {Baron}, {Pedretti},
  {Thureau}, \& {Webster}}]{Che_2012_07_0}
{Che}, X., {Monnier}, J.~D., {Kraus}, S., {et~al.} 2012, in Society of
  Photo-Optical Instrumentation Engineers (SPIE) Conference Series, ed.
  F.~{Delplancke}, J.~K. {Rajagopal}, \& F.~{Malbet}, Vol. 8445

\bibitem[{{Che} {et~al.}(2010){Che}, {Monnier}, \& {Webster}}]{Che_2010_07_0}
{Che}, X., {Monnier}, J.~D., \& {Webster}, S. 2010, in Society of Photo-Optical
  Instrumentation Engineers (SPIE) Conference Series, ed. W.~C. {Danchi},
  F.~{Delplancke}, \& J.~K. {Rajagopal}, Vol. 7734, 77342V--77342V--9

\bibitem[{{Claret} \& {Bloemen}(2011)}]{Claret_2011_05_0}
{Claret}, A. \& {Bloemen}, S. 2011, \aap, 529, A75

\bibitem[{{Cutri} {et~al.}(2003){Cutri}, {Skrutskie}, {van Dyk}, {Beichman},
  {Carpenter}, {Chester}, {Cambresy}, {Evans}, {Fowler}, {Gizis}, {Howard},
  {Huchra}, {Jarrett}, {Kopan}, {Kirkpatrick}, {Light}, {Marsh}, {McCallon},
  {Schneider}, {Stiening}, {Sykes}, {Weinberg}, {Wheaton}, {Wheelock}, \&
  {Zacarias}}]{Cutri_2003_03_0}
{Cutri}, R.~M., {Skrutskie}, M.~F., {van Dyk}, S., {et~al.} 2003, VizieR Online
  Data Catalog, 2246, 0

\bibitem[{{Evans}(2000)}]{Evans_2000_06_0}
{Evans}, N.~R. 2000, \aj, 119, 3050

\bibitem[{{Evans} {et~al.}(2006){Evans}, {Franz}, {Massa}, {Mason}, {Walker},
  \& {Karovska}}]{Evans_2006_11_0}
{Evans}, N.~R., {Franz}, O., {Massa}, D., {et~al.} 2006, \pasp, 118, 1545

\bibitem[{{Evans} {et~al.}(2008){Evans}, {Schaefer}, {Bond}, {Bono},
  {Karovska}, {Nelan}, {Sasselov}, \& {Mason}}]{Evans_2008_09_0}
{Evans}, N.~R., {Schaefer}, G.~H., {Bond}, H.~E., {et~al.} 2008, \aj, 136, 1137

\bibitem[{{Fernie}(1969)}]{Fernie_1969_12_0}
{Fernie}, J.~D. 1969, \pasp, 81, 707

\bibitem[{{Fouqu{\'e}} \& {Gieren}(1997)}]{Fouque_1997_04_0}
{Fouqu{\'e}}, P. \& {Gieren}, W.~P. 1997, \aap, 320, 799

\bibitem[{{Fouqu{\'e}} {et~al.}(2003){Fouqu{\'e}}, {Storm}, \&
  {Gieren}}]{Fouque_2003__0}
{Fouqu{\'e}}, P., {Storm}, J., \& {Gieren}, W. 2003, in {Lecture Notes in
  Physics}, Vol. 635, Stellar Candles for the Extragalactic Distance Scale, ed.
  D.~{Alloin} \& W.~{Gieren} (Berlin Springer), 21

\bibitem[{{Gieren} {et~al.}(1998){Gieren}, {Fouque}, \&
  {Gomez}}]{Gieren_1998_03_0}
{Gieren}, W.~P., {Fouque}, P., \& {Gomez}, M. 1998, \apj, 496, 17

\bibitem[{{Hanbury Brown} {et~al.}(1974){Hanbury Brown}, {Davis}, {Lake}, \&
  {Thompson}}]{Hanbury-Brown_1974_06_0}
{Hanbury Brown}, R., {Davis}, J., {Lake}, R.~J.~W., \& {Thompson}, R.~J. 1974,
  \mnras, 167, 475

\bibitem[{{Hartkopf} \& {McAlister}(1984)}]{Hartkopf_1984_01_0}
{Hartkopf}, W.~I. \& {McAlister}, H.~A. 1984, \pasp, 96, 105

\bibitem[{{Keller}(2008)}]{Keller_2008_04_0}
{Keller}, S.~C. 2008, \apj, 483

\bibitem[{{Kervella} {et~al.}(2004{\natexlab{a}}){Kervella}, {Bersier},
  {Mourard}, {Nardetto}, {Fouqu{\'e}}, \& {Coud{\'e} du
  Foresto}}]{Kervella_2004_12_0}
{Kervella}, P., {Bersier}, D., {Mourard}, D., {et~al.} 2004{\natexlab{a}},
  \aap, 428, 587

\bibitem[{{Kervella} {et~al.}(2004{\natexlab{b}}){Kervella}, {Th{\'e}venin},
  {Di Folco}, \& {S{\'e}gransan}}]{Kervella_2004_10_0}
{Kervella}, P., {Th{\'e}venin}, F., {Di Folco}, E., \& {S{\'e}gransan}, D.
  2004{\natexlab{b}}, \aap, 426, 297

\bibitem[{{Kiss} \& {Vink{\'o}}(2000)}]{Kiss_2000_05_0}
{Kiss}, L.~L. \& {Vink{\'o}}, J. 2000, \mnras, 314, 420

\bibitem[{{Klagyivik} \& {Szabados}(2009)}]{Klagyivik_2009_09_0}
{Klagyivik}, P. \& {Szabados}, L. 2009, \aap, 504, 959

\bibitem[{{Kovtyukh} {et~al.}(2008){Kovtyukh}, {Soubiran}, {Luck}, {Turner},
  {Belik}, {Andrievsky}, \& {Chekhonadskikh}}]{Kovtyukh_2008_09_0}
{Kovtyukh}, V.~V., {Soubiran}, C., {Luck}, R.~E., {et~al.} 2008, \mnras, 389,
  1336

\bibitem[{{Luck} {et~al.}(2008){Luck}, {Andrievsky}, {Fokin}, \&
  {Kovtyukh}}]{Luck_2008_07_0}
{Luck}, R.~E., {Andrievsky}, S.~M., {Fokin}, A., \& {Kovtyukh}, V.~V. 2008,
  \aj, 136, 98

\bibitem[{{McAlister}(1978)}]{McAlister_1978_06_0}
{McAlister}, H.~A. 1978, \pasp, 90, 288

\bibitem[{{Millis}(1969)}]{Millis_1969__0}
{Millis}, R.~L. 1969, Lowell Observatory Bulletin, 7, 113

\bibitem[{{Monnier} {et~al.}(2010){Monnier}, {Anderson}, {Baron}, {Berger},
  {Che}, {Eckhause}, {Kraus}, {Pedretti}, {Thureau}, {Millan-Gabet}, {Ten
  Brummelaar}, {Irwin}, \& {Zhao}}]{Monnier_2010_07_0}
{Monnier}, J.~D., {Anderson}, M., {Baron}, F., {et~al.} 2010, in Society of
  Photo-Optical Instrumentation Engineers (SPIE) Conference Series, ed. W.~C.
  {Danchi}, F.~{Delplancke}, \& J.~K. {Rajagopal}, Vol. 7734,
  77340G--77340G--12

\bibitem[{{Monnier} {et~al.}(2004){Monnier}, {Berger}, {Millan-Gabet}, \& {ten
  Brummelaar}}]{Monnier_2004_10_0}
{Monnier}, J.~D., {Berger}, J.-P., {Millan-Gabet}, R., \& {ten Brummelaar},
  T.~A. 2004, in Society of Photo-Optical Instrumentation Engineers (SPIE)
  Conference Series, ed. {W.~A.~Traub}, Vol. 5491, 1370

\bibitem[{{Monnier} {et~al.}(2012){Monnier}, {Che}, {Zhao}, {Ekstr{\"o}m},
  {Maestro}, {Aufdenberg}, {Baron}, {Georgy}, {Kraus}, {McAlister}, {Pedretti},
  {Ridgway}, {Sturmann}, {Sturmann}, {ten Brummelaar}, {Thureau}, {Turner}, \&
  {Tuthill}}]{Monnier_2012_12_0}
{Monnier}, J.~D., {Che}, X., {Zhao}, M., {et~al.} 2012, \apjl, 761, L3

\bibitem[{{Monnier} {et~al.}(2007){Monnier}, {Zhao}, {Pedretti}, {Thureau},
  {Ireland}, {Muirhead}, {Berger}, {Millan-Gabet}, {Van Belle}, {ten
  Brummelaar}, {McAlister}, {Ridgway}, {Turner}, {Sturmann}, {Sturmann}, \&
  {Berger}}]{Monnier_2007_07_0}
{Monnier}, J.~D., {Zhao}, M., {Pedretti}, E., {et~al.} 2007, Science, 317, 342

\bibitem[{{Moskalik} \& {Gorynya}(2005)}]{Moskalik_2005_06_0}
{Moskalik}, P. \& {Gorynya}, N.~A. 2005, Acta Astronomica, 55, 247

\bibitem[{{Neilson} {et~al.}(2011){Neilson}, {Cantiello}, \&
  {Langer}}]{Neilson_2011_05_0}
{Neilson}, H.~R., {Cantiello}, M., \& {Langer}, N. 2011, \aap, 529, L9

\bibitem[{{Pietrzy{\'n}ski} {et~al.}(2010){Pietrzy{\'n}ski}, {Thompson},
  {Gieren}, {Graczyk}, {Bono}, {Udalski}, {Soszy{\'n}ski}, {Minniti}, \&
  {Pilecki}}]{Pietrzynski_2010_11_0}
{Pietrzy{\'n}ski}, G., {Thompson}, I.~B., {Gieren}, W., {et~al.} 2010, \nat,
  468, 542

\bibitem[{{Pietrzy{\'n}ski} {et~al.}(2011){Pietrzy{\'n}ski}, {Thompson},
  {Graczyk}, {Gieren}, {Pilecki}, {Udalski}, {Soszynski}, {Bono}, {Konorski},
  {Nardetto}, \& {Storm}}]{Pietrzynski_2011_12_0}
{Pietrzy{\'n}ski}, G., {Thompson}, I.~B., {Graczyk}, D., {et~al.} 2011, \apjl,
  742, L20

\bibitem[{{Prada Moroni} {et~al.}(2012){Prada Moroni}, {Gennaro}, {Bono},
  {Pietrzy{\'n}ski}, {Gieren}, {Pilecki}, {Graczyk}, \&
  {Thompson}}]{Prada-Moroni_2012_04_0}
{Prada Moroni}, P.~G., {Gennaro}, M., {Bono}, G., {et~al.} 2012, \apj, 749, 108

\bibitem[{{Samus} {et~al.}(2009){Samus}, {Durlevich},
  {et~al.}}]{Samus_2009_01_0}
{Samus}, N.~N., {Durlevich}, O.~V., {et~al.} 2009, VizieR Online Data Catalog:
  B/gcvs. Originally published in: Institute of Astronomy of Russian Academy of
  Sciences and Sternberg State Astronomical Institute of the Moscow State
  University, 1, 2025

\bibitem[{{Sandage} \& {Tammann}(2006)}]{Sandage_2006_09_0}
{Sandage}, A. \& {Tammann}, G.~A. 2006, \araa, 44, 93

\bibitem[{{Scardia} {et~al.}(2008){Scardia}, {Prieur}, {Pansecchi}, {Argyle},
  {Sala}, {Basso}, {Ghigo}, {Koechlin}, \& {Aristidi}}]{Scardia_2008_01_0}
{Scardia}, M., {Prieur}, J.-L., {Pansecchi}, L., {et~al.} 2008, Astronomische
  Nachrichten, 329, 54

\bibitem[{{Storm} {et~al.}(2011){Storm}, {Gieren}, {Fouqu{\'e}}, {Barnes},
  {Pietrzy{\'n}ski}, {Nardetto}, {Weber}, {Granzer}, \&
  {Strassmeier}}]{Storm_2011_10_0}
{Storm}, J., {Gieren}, W., {Fouqu{\'e}}, P., {et~al.} 2011, \aap, 534, A94

\bibitem[{{Szabados}(1991)}]{Szabados_1991_01_0}
{Szabados}, L. 1991, Commun. of the Konkoly Observatory Hungary, 96, 123

\bibitem[{{Szabados}(2003)}]{Szabados_2003__0}
{Szabados}, L. 2003, in Astronomical Society of the Pacific Conference Series,
  Vol. 298, GAIA Spectroscopy: Science and Technology, ed. {U.~Munari}, 237

\bibitem[{{Tallon-Bosc} {et~al.}(2008){Tallon-Bosc}, {Tallon}, {Thi{\'e}baut},
  {B{\'e}chet}, {Mella}, {Lafrasse}, {Chesneau}, {Domiciano de Souza},
  {Duvert}, {Mourard}, {Petrov}, \& {Vannier}}]{Tallon-Bosc_2008_07_0}
{Tallon-Bosc}, I., {Tallon}, M., {Thi{\'e}baut}, E., {et~al.} 2008, in Society
  of Photo-Optical Instrumentation Engineers (SPIE) Conference Series, Vol.
  7013, Society of Photo-Optical Instrumentation Engineers (SPIE) Conference
  Series

\bibitem[{{ten Brummelaar} {et~al.}(2005){ten Brummelaar}, {McAlister},
  {Ridgway}, {Bagnuolo}, {Turner}, {Sturmann}, {Sturmann}, {Berger}, {Ogden},
  {Cadman}, {Hartkopf}, {Hopper}, \& {Shure}}]{ten-Brummelaar_2005_07_0}
{ten Brummelaar}, T.~A., {McAlister}, H.~A., {Ridgway}, S.~T., {et~al.} 2005,
  \apj, 628, 453

\bibitem[{{van Leeuwen}(2007)}]{van-Leeuwen_2007_11_0}
{van Leeuwen}, F. 2007, \aap, 474, 653

\bibitem[{{Wright} \& {Howard}(2009)}]{Wright_2009_05_0}
{Wright}, J.~T. \& {Howard}, A.~W. 2009, \apjs, 182, 205

\end{thebibliography}

\end{document}